\documentclass[conference,letterpaper]{IEEEtran}
\usepackage{amssymb}
\usepackage{pgfplots}
\usepackage{graphicx}
\usepackage{comment}

\usepackage[utf8]{inputenc} 
\usepackage[T1]{fontenc}
\usepackage{ifthen}
\usepackage{cite}
\usepackage[cmex10]{amsmath} 
\newcommand{\E}[1]{\mathrm{E}\left[#1\right]}

\begin{document}
\title{Feedback Gains for Gaussian Massive Multiple-Access Channels}

\author{%
  \IEEEauthorblockN{Gerhard~Kramer}
  \IEEEauthorblockA{Institute for Communications Engineering\\
                    Technical University of Munich\\
                    D-80333 Munich, Germany}
}

\maketitle

\begin{abstract}
Feedback is shown to increase the sum-rate capacity of K-user Gaussian multiple-access
channels  by at most a factor of approximately 1.54, improving Thomas' doubling bound (1987).
The new bound is the best possible in the sense that it can be approached as closely as desired
for a massive number of users. Moreover, feedback provides unbounded power gain in K for a
fixed transmit power per user.
\end{abstract}

\section{Introduction}
\label{sec:introduction}
The $K$-user ($K\ge2$) complex-alphabet Gaussian multiple-access channel (MAC) has output
\begin{align}
   Y = Z + \sum_{k=1}^K X_k
\end{align}
where the $X_k$, $k=1,\dots,K$, are complex channel inputs and $Z$ is complex, circularly-symmetric, Gaussian
noise with unit variance. Consider the average block power constraints
\begin{align}
   \frac{1}{n} \sum_{i=1}^n \E{\left| X_{k,i} \right|^2} \le P_k, \quad k=1,\dots,K
\end{align}
where $\E{.}$ denotes expectation, $X_{k,i}$ represents the channel input of user $k$ at time
$i$, $i=1,\dots,n$, and $P_k>0$ for all $k$.  We remark that real-alphabet channels can be treated as
complex-alphabet channels by grouping the real symbols into pairs that are treated as complex numbers.

The sum-rate capacity of the MAC \emph{without} feedback in nats per channel use is known
to be (see~\cite{Thomas-IT87}) 
\begin{align} \label{eq:C}
   C(P_1,\dots,P_K) = \ln\left( 1 + K P \right)
\end{align}
where $P=\frac{1}{K}\sum_{k=1}^K P_k$ is the average transmitter power.
The sum-rate capacity \emph{with} feedback is unknown in general, even for the most informative 
causal feedback, i.e., the past channel outputs $Y_1,\dots,Y_{i-1}$ at time $i$.
A classic paper by Thomas~\cite{Thomas-IT87} shows that feedback can at most double the
sum-rate capacity. The result follows by treating symmetric power constraints, $P_k=P$ for all $k$,
and then generalizing via a concavity
argument~\cite[Sec.~V]{Thomas-IT87}. More precisely, the feedback sum-rate capacity
for \emph{general} power constraints satisfies
\begin{align} 
   & C_{\text{FB}}(P_1,\dots,P_K) \le C_{\text{FB}}(P,\dots,P) \nonumber \\
   & \qquad \le 2 C(P,\dots,P) = 2 C(P_1,\dots,P_K).
   \label{eq:Thomas}
\end{align}

The sum-rate capacity is known for some interesting special cases.
Ozarow~\cite{Ozarow-IT84} determined the capacity region for $K=2$
and Kramer~\cite{Kramer-IT02} found the sum-rate capacity for $K\ge 3$ under
symmetric power constraints, $P_k=P$ for all $k$, and if $P$ is beyond some threshold.
For both these cases, the capacity matches a cut-set bound.
Recent work by Sula et al.~\cite{Sula-Gastpar-Kramer-IT20} builds
on~\cite{Kramer-IT02,Hekstra-Willems-IT89,Kramer-IT03,Kramer-Gastpar-ITW06,Ard-Wigger-Kim-Javidi-IT12}
and shows that
\begin{align} \label{eq:CFB}
   C_{\text{FB}}(P,\dots,P) = \ln\left( 1 + K P \lambda^* \right)
\end{align}
where $\lambda^*$ is the unique real number in the interval $[1,K]$ that satisfies
the dependence-balance bound (see~\cite{Hekstra-Willems-IT89,Kramer-IT03,Kramer-Gastpar-ITW06})
\begin{align} \label{eq:DB}
   \frac{1}{K} \ln\left(1 + K P \lambda\right) \le \frac{1}{K-1} \ln\left(1 + (K-\lambda) P \lambda \right)
\end{align}
with equality. The coding theorem was established in~\cite{Kramer-IT02}
and the converse is proved in~\cite{Sula-Gastpar-Kramer-IT20}.

This paper studies the power gain factor $\lambda^*$ in \eqref{eq:CFB} and the capacity gain factor
\begin{align} \label{eq:F}
   F(\pi) = \frac{\ln\left(1 + \pi \lambda^* \right)}{\ln\left(1 + \pi \right)}
\end{align}
where $\pi=K P$ is the total transmit power.
Sec.~\ref{sec:feedback-gains} develops bounds on these factors and gives numerical results.
For example, we improve the second inequality in \eqref{eq:Thomas} to
\begin{align} \label{eq:CFB2}
   C_{\text{FB}}(P,\dots,P) \le 1.54 \cdot C(P,\dots,P)
\end{align}
which also improves \eqref{eq:Thomas} for general power constraints.
Sec.~\ref{sec:conclusions} concludes the paper.

\section{Feedback Gains}
\label{sec:feedback-gains}
\begin{figure}[t]
      \centering
      \includegraphics[width=0.95\columnwidth]{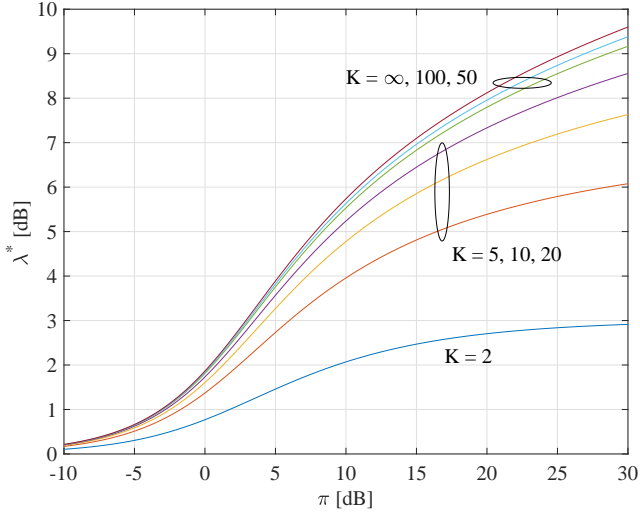}
      \caption{Power gain factor $\lambda^*$ vs. $\pi$.}
      \label{fig:Pfactor}
\end{figure}

\begin{figure}[t]
      \centering
      \includegraphics[width=0.95\columnwidth]{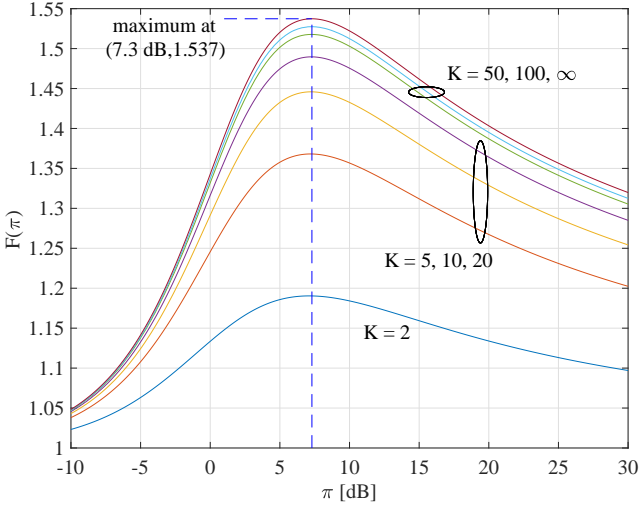}
      \caption{Capacity gain factor $F(\pi)$ vs. $\pi$.}
      \label{fig:Cfactor}
\end{figure}

Fig.~\ref{fig:Pfactor} and Fig.~\ref{fig:Cfactor} plot $\lambda^*$ and $F(\pi)$ against $\pi$ in decibels
for selected $K$. We prove several properties of these curves. Consider the simple bounds
\begin{align}
  \frac{x}{1+x} \le \ln(1+x) \le x
  \label{eq:ln-bounds}
\end{align}
for $x>-1$ with equality on both sides if and only if $x=0$.
We follow the proof steps of~\cite[Thm.~3]{Kramer-IT02}
and write \eqref{eq:DB} with equality as in~\cite[Eq.~(70)]{Kramer-IT02}:
\begin{align}
   \ln\left(1 + \pi \lambda\right)
   = K \ln\left(1 + \frac{P \lambda^2}{1 + (K-\lambda) P \lambda}  \right).
   \label{eq:DBnew}
\end{align}
Now apply \eqref{eq:ln-bounds} to bound
\begin{align}
   \frac{\pi \lambda^2}{1 + \pi \lambda}
   \le \ln\left(1 + \pi \lambda\right)
   \le \frac{\pi \lambda^2}{1 + (K - \lambda) P \lambda}
   \label{eq:bounds1}
\end{align}
with equality on both sides of \eqref{eq:bounds1} if and only if $P=0$.
The right-hand side (RHS) of \eqref{eq:bounds1} is less than
$K\lambda/(K-\lambda)$ and rearranging terms we obtain (cf.~\cite[Eq.~(71)]{Kramer-IT02})
\begin{align} 
  \frac{K \ln\left(1 + \pi \lambda \right)}{K + \ln\left(1 + \pi \lambda \right)}
  < \lambda \le f(\pi,\lambda)
  \label{eq:bounds2}
\end{align}
where
\begin{align} 
  f(\pi,\lambda) = \left( 1 + \frac{1}{\pi \lambda} \right) \ln\left(1 + \pi \lambda \right).
  \label{eq:f}
\end{align}


\subsection*{Properties of $\lambda^*$}
Fig.~\ref{fig:Pfactor} illustrates that $\lambda^*$ has the following properties:
\begin{enumerate}
\item \label{list:prop-max} $\lambda^*$ is bounded and maximized by $K \rightarrow \infty$ for fixed $\pi$; 
\item \label{list:prop-approach-K} $\lambda^* \rightarrow K$ for fixed $K$ and $P \rightarrow \infty$; 
\item \label{list:prop-approach-finite} $\lambda^* \rightarrow \infty$ for $K \rightarrow \infty$ and $\pi \rightarrow \infty$.
\end{enumerate}
To prove Property~\ref{list:prop-max}, fix $\pi$ and use the left-hand side (LHS)
of \eqref{eq:ln-bounds} to show that
\begin{align}
   f(\pi,1) > 1, \quad 0 < \frac{\partial f(\pi,\lambda)}{\partial \lambda} < \frac{\pi}{1+\pi \lambda} \le 1
   \label{eq:f-bounds}
\end{align}
for $P>0$. The bounds \eqref{eq:f-bounds} imply that the equation
\begin{align}
   \lambda = f(\pi,\lambda)
   \label{eq:Kinf}
\end{align}
has a unique and finite solution for $\lambda\ge1$. Moreover, equality holds on both sides
of \eqref{eq:bounds1}, and on the RHS of \eqref{eq:bounds2}, for $P\rightarrow0$. 
Thus, the maximal $\lambda$ is attained by $K\rightarrow\infty$, i.e., the uppermost
curves in Fig.~\ref{fig:Pfactor} and Fig.~\ref{fig:Cfactor} are defined by \eqref{eq:Kinf}.

Properties~\ref{list:prop-approach-K} and~\ref{list:prop-approach-finite} follow by
the LHS of \eqref{eq:bounds2} and $\lambda \ge 1$.
The bounds \eqref{eq:bounds2} also imply that $\log \lambda$ scales as $\log \log K$ for fixed $P$, see~\cite[Thm.~3]{Kramer-IT02}.

\subsection*{Properties of $F(\pi)$}
Fig.~\ref{fig:Cfactor} shows that $F(\pi)$ has the following properties:
\begin{enumerate} \setcounter{enumi}{3}
\item \label{list:prop-F-small} $F(\pi) \rightarrow 1$ for $\pi \rightarrow 0$ or $\pi \rightarrow \infty$;
\item \label{list:prop-F-peak} the maximal capacity gain factor is $F(\pi)\approx1.537$ at $\pi\approx 5.38$ (or  $\pi\approx 7.3$ dB); 
\item \label{list:prop-F-bound} $F(\pi)$ is not too sensitive to $\pi$ for $\pi\ge 5$ dB.
\end{enumerate}
To prove Property~\ref{list:prop-F-small} for small $\pi$, apply the RHS of \eqref{eq:ln-bounds} 
to the RHS of~\eqref{eq:bounds2} and rearrange terms to obtain
\begin{align}
   (1-\pi)\lambda \le 1
    \label{eq:pi-lambda-bound}
\end{align}
with equality if and only if $\pi=0$. Now apply \eqref{eq:ln-bounds} and \eqref{eq:pi-lambda-bound}
to \eqref{eq:F} to bound
\begin{align}
  F(\pi) \le (1+\pi) \lambda \le \frac{1+\pi}{1-\pi}, \; \text{ if } 0 \le \pi <1 .
  \label{eq:F-bound-1}
\end{align}
We thus have $\lambda\rightarrow1$ and $F(\pi)\rightarrow1$ as $\pi\rightarrow0$.
Next, for large $\pi$ we have $f(\pi,\lambda) \approx \ln(1+\pi \lambda)$ and
in the limit of large $K$ and $\pi$ we have $\lambda=\ln(1+\pi\lambda)$. We thus have
\begin{align}
   \lim_{\pi \rightarrow \infty} F(\pi) 
   \le \lim_{\lambda \rightarrow \infty} \frac{\lambda}{\ln\left( 1 + \left( e^\lambda-1 \right) / \lambda \right)} = 1.
\end{align}

Property~\ref{list:prop-F-peak} is trickier to prove because the uppermost curve
in Fig.~\ref{fig:Cfactor} is defined by the implicit equation \eqref{eq:Kinf}. We thus
numerically evaluate $F(\pi)$ for a range of $\pi$ and derive bounds for the other $\pi$.
Fig.~\ref{fig:Cfactor} shows that the uppermost curve is maximally $F(\pi)\approx1.537$
at $\pi\approx 7.3$ dB in the interval $-10\textrm{ dB} \le \pi \le 30 \textrm{ dB}$.
The Appendix treats $\pi \le -10$ dB and $\pi \ge 30$ dB and shows that $F(\pi)\le 1.321$ for these $\pi$.
In fact, numerical calculations show that all curves in Fig.~\ref{fig:Cfactor} are unimodal,
i.e., all curves increase with $\pi$ from $F(0)=1$ up to a peak and then decrease with $\pi$ back to $F(\infty)=1$. 

Property~\ref{list:prop-F-peak} strengthens Thomas' factor-of-two bound.
Moreover, the new bound is the best possible in the sense that it can be approached as closely
as desired as $K\rightarrow\infty$. For example, for $K=100$ and $P=0$ dB we have 
$\lambda^* \approx 8$ dB and $F(\pi) \approx 1.4$. Also, for finite $K$ we have
tighter bounds than for massive $K$, e.g., for $K=10$ the maximal capacity gain factor is
$F(\pi)\approx1.446$ at $\pi\approx 7.2$ dB.

\section{Conclusions}
\label{sec:conclusions}
Feedback was shown to increase the sum-rate capacity of $K$-user Gaussian MACs by at most
a factor of approximately 1.54. The new bound can be approached for a massive number of users
at a total transmit power of $\pi \approx 7.3$ dB.
Also, feedback provides unbounded power gain in $K$ for a fixed transmit power per user.
Note that we have studied memoryless channels and it is interesting to consider channels
with memory, see~\cite{Kim-IT10} for references on single-user channels
and~\cite{Pombra-Cover-IT94,Ordentlich96} for MACs.

Finally, we remark that channel-output feedback is usually considered unrealistic for classic
communications applications because of its high precision that requires high rate.
However, future radio standards such as 6G also target control applications for mechanical
systems that react slowly as compared to wireless communication speeds. 
If we view the uplink as being mechanical and slow, then the wireless downlink
could have high rate. A second interesting application is internet-of-things (IoT) systems
with large $K$ and low uplink communication rates but where the downlink may have
high capacity.

\appendix
Implicit differentiation of \eqref{eq:Kinf} gives
\begin{align}
   \frac{d\lambda}{d\pi} = \frac{\lambda}{\pi}
   \cdot \frac{(\pi-1)\lambda+1}{\pi\lambda(\lambda-1)+2\lambda-1}
\end{align}
which is positive for both $\pi \ge 1$ and $0<\pi<1$ via \eqref{eq:pi-lambda-bound}.
The power gain factor $\lambda$ thus increases with $\pi$, i.e., the uppermost curve in
Fig.~\ref{fig:Pfactor} is increasing. In fact, all curves in Fig.~\ref{fig:Pfactor} are increasing.

Consider now the range $\pi \le -10$  dB. The RHS of \eqref{eq:F-bound-1} increases with $\pi$ and
we thus have  $F(\pi) \le 11/9\approx1.222$. We see that $F(\pi) \le 1.321$, as claimed, and
the bound is loose since $F(0.1) \approx 1.048$.

Consider next the range $\pi \ge 30$ dB. Equation \eqref{eq:Kinf} gives
\begin{align}
   & \lambda \ge \ln(1+\pi\lambda) \\
   & \pi = \left. \left( \exp\left( \frac{\pi\lambda^2}{1+\pi \lambda} \right) - 1 \right) \right/ \lambda
\end{align}
and therefore
\begin{align}
  F(\pi) & \le \frac{\lambda}{\ln\left( \exp\left( \frac{\pi\lambda^2}{1+\pi \lambda} \right) + \lambda - 1 \right) - \ln \lambda}
  \nonumber \\
  & \le \left[ \frac{\pi\lambda}{1+\pi \lambda} - (\ln \lambda)/\lambda \right]^{-1}.
  \label{eq:F-bound-2}
\end{align}
The term in square brackets in \eqref{eq:F-bound-2}
increases with $\pi$ if $\lambda>e$ because:
\begin{itemize}
\item $\lambda$ increases with $\pi$;
\item $\pi \lambda/(1+\pi \lambda)$ increases with $\pi \lambda$;
\item $(\ln \lambda)/\lambda$ decreases in $\lambda$ for $\lambda>e$.
\end{itemize}
But for $\pi=30$ dB we have $\lambda \approx 9.119>e$, see Fig.~\ref{fig:Pfactor},
and \eqref{eq:F-bound-2} gives $F(\pi) \le 1.321$, as claimed.
We compute $F(1000) \approx 1.320$ so the bound is almost tight.

\section*{Acknowledgement}
This work was supported by the German Research Foundation (DFG) under Grant {KR 3517/9-1}.



\end{document}